\documentclass[journal]{IEEEtran}

\usepackage{cite}
\usepackage{amsmath,amssymb,amsfonts}
\usepackage{algorithmic}
\usepackage{graphicx}
\usepackage{textcomp}
\usepackage{xcolor}
\usepackage{url}
\usepackage{svg}
\usepackage{adjustbox}
\usepackage{array}
\usepackage{booktabs}

\newcommand{\mycomment}[1]{}

\def\BibTeX{{\rm B\kern-.05em{\sc i\kern-.025em b}\kern-.08em
    T\kern-.1667em\lower.7ex\hbox{E}\kern-.125emX}}
\begin{document}

\title{Cuff-less Arterial Blood Pressure Waveform Synthesis from Single-site PPG using Transformer \& Frequency-domain Learning 
{\footnotesize \textsuperscript{}}
}

\author{
\IEEEauthorblockN{
Muhammad Wasim Nawaz\IEEEauthorrefmark{1}, Muhammad Ahmad Tahir\IEEEauthorrefmark{2}, Ahsan Mehmood\IEEEauthorrefmark{2}, Muhammad Mahboob Ur Rahman\IEEEauthorrefmark{2}, \\
Kashif Riaz\IEEEauthorrefmark{2}, Qammer H. Abbasi\IEEEauthorrefmark{3},\IEEEauthorrefmark{4} } \\
\thanks{This work is supported in part by UK Engineering and Physical Sciences Research Council (EPSRC) under the grants: EP/T021063/1 and EP/T021020/1.}
\IEEEauthorblockA{
\IEEEauthorrefmark{1}Computer engineering department, University of Lahore, Lahore, Pakistan\\
\IEEEauthorrefmark{2} Electrical engineering department, Information Technology University, Lahore 54000, Pakistan\\ 
\IEEEauthorrefmark{3}James Watt School of Engineering, University of Glasgow, Glasgow, G12 8QQ, UK\\
\IEEEauthorrefmark{4}Artificial Intelligence Research Centre, Ajman University, Ajman, UAE\\
\IEEEauthorrefmark{1}muhammad.wasim@dce.uol.edu.pk, \IEEEauthorrefmark{2}mahboob.rahman@itu.edu.pk, \IEEEauthorrefmark{3}Qammer.Abbasi@glasgow.ac.uk 
}
}

\maketitle
\newpage

\begin{abstract}

We develop and evaluate two novel purpose-built deep learning (DL) models for synthesis of the arterial blood pressure (ABP) waveform in a cuff-less manner, using a single-site photoplethysmography (PPG) signal. We train and evaluate our DL models on the data of 209 subjects from the public UCI dataset on cuff-less blood pressure (CLBP) estimation. Our transformer model consists of an encoder-decoder pair that incorporates positional encoding, multi-head attention, layer normalization, and dropout techniques for ABP waveform synthesis. Secondly, under our frequency-domain (FD) learning approach, we first obtain the discrete cosine transform (DCT) coefficients of the PPG and ABP signals, and then learn a linear/non-linear (L/NL) regression between them. The transformer model (FD L/NL model) synthesizes the ABP waveform with a mean absolute error (MAE) of 3.01 (4.23). 
Further, the synthesis of ABP waveform also allows us to estimate the systolic blood pressure (SBP) and diastolic blood pressure (DBP) values.
To this end, the transformer model reports an MAE of 3.77 mmHg and 2.69 mmHg, for SBP and DBP, respectively. On the other hand, the FD L/NL method reports an MAE of 4.37 mmHg and 3.91 mmHg, for SBP and DBP, respectively. Both methods fulfill the AAMI criterion. As for the BHS criterion, our transformer model (FD L/NL regression model) achieves grade A (grade B). 

\end{abstract}


\section{Introduction}

Hypertension (high blood pressure) and hypotension (low blood pressure) are two common blood pressure (BP) irregularities that are of great clinical significance. According to the World Health Organization (WHO), approximately 1.28 billion people aged 30-79 years suffer from hypertension, with 46\% unaware of it and only 42\% receiving treatment \cite{who2023hypertension}. Further, hypertension is more prevalent in low and middle income countries \cite{nugroho2022comparison}. Hypertension (also known as the silent killer) is a major risk factor for a range of cardiovascular diseases, stroke, chronic kidney disease and pre-mature death \cite{zhou2021global}. Hypertension, if not treated, could elevate both the systolic BP (SBP) and diastolic BP (DBP) from normal range (below 120/below 80) to stage 1 hypertension (130-139/80-89), to stage 2 hypertension (above 140/above 90), to hypertensive crisis (above 180/above 120), leading to health emergency necessitating immediate hospitalization. 
Hypotension, though less frequent than hypertension, is also widespread as approximately (10-20)\% of people of age over 65 years experience hypotension. Hypotension occurs when blood pressure falls below 90/60, which could lead to dizziness and nausea. Hypotension becomes perilous when BP falls below 60/45, causing blurriness, confusion, coma, or even death \cite{saedon2020prevalence}. Thus, regular and frequent monitoring of blood pressure by the elderly is of utmost importance for early diagnosis and efficient management of hypertension and hypotension \cite{tucker2017self}. 


The existing clinical-grade methods for blood pressure estimation (including ambulatory devices) could be classified into two categories: i) invasive methods, ii) non-invasive but cuff-based methods. Invasive methods are the gold standard methods which measure the centralized aortic blood pressure by inserting catheters in the aorta (main artery near the heart) \cite{saugel2020measure}. Need not to say that this method is painful, prone to infections, and requires trained professionals in a clinical setting. Next, non-invasive but cuff-based methods work by inflating the cuff around the arm and rely upon manual auscultation and Sphygmomanometers for peripheral blood pressure estimation. These methods, though highly accurate, suffer from a number of limitations, e.g., inconvenience of regular monitoring, requirement of trained personnel, use of toxic mercury by Sphygmomanometers, etc. \cite{pandit2020cuffless, kachuee2016cuffless, 8440703}. Further, the need for frequent cuff inflation and deflation makes continuous BP measurement impractical, especially for individuals with weaker arms, or the elderly \cite{1403635}. Oscillometric devices do not require manual auscultation, but they have relatively lower accuracy as they rely on mean arterial blood pressure (MABP) for estimating both SBP and DBP. 

The limitations of the existing invasive methods and cuff-based methods have thus prompted the researchers to design novel cuff-less blood pressure (CLBP) estimation methods which are meant to be user-friendly, reliable, and support continuous and in-situ BP measurement \cite{pandit2020cuffless, pandit_lores_batlle_2020, sola2019handbook, 6346359}. Additionally, advancements in wearable technology (e.g., smart watches, smart bands, fitness trackers, internet of medical things, flexible electronic tattoos, etc.) and assistive gadgets (e.g., smartphones, airpods, etc.) have further accelerated the development of the CLBP estimation methods \cite{korean_position_paper}, \cite{8650069}. CLBP estimation methods either rely upon bio-signals such as photoplethysmography (PPG), electrocardiography (ECG), phonocardiography (PCG) acquired non-invasively from one or more body sites \cite{sola2019handbook}, or innovative sensor technology, e.g., tonometry \cite{4353109}.




{\bf Regulatory criteria for CLBP estimation methods:}
Over time, two regulatory bodies have developed standards to grade the performance of the emerging CLBP estimation methods: 1) association for the advancement of medical instrumentation (AAMI), and 2) British hypertension society (BHS)\footnote{Additionally, IEEE has also developed 1708-2014 (and 1708a-2019) standard for wearable cuffless blood pressure measuring devices \cite{6846264}.}. 
\begin{itemize}
    \item The AAMI requires a dataset consisting of at least 85 individuals having low, high, and normal blood pressure. Further, according to AAMI, the mean error (and standard deviation of the error) of a CLBP estimation method should be within 5 mmHg (and 8 mmHg) \cite{ansi_aami_sp10-192}. 
    \item The BHS assigns one of the three letter grades (A, B, C) to each CLBP estimation method. If the model error is less than or equal to 5 mmHg 60\% of time, 10 mmHg 85\% of time, and 15 mmHg 95\% of time, then the model is given grade A. Similarly, if the model error is less than 5 mmHg 50\% of time, 10 mmHg 75\% of time, and 15 mmHg 90\% of time, then the model is given grade B. Finally, if the model error is less than 5 mmHg 40\% of time, 10 mmHg 65\% of time, and 15 mmHg 85\% of time, then the model is given grade C \cite{bhs1993}.
\end{itemize}

{\bf Contributions.} 
Inline with the recent works on CLBP estimation, we propose two novel purpose-built deep learning (DL) models for synthesis of the arterial blood pressure (ABP) waveform in a cuff-less manner, using a single-site PPG signal. We utilize the public UCI dataset on CLBP estimation to train and evaluate our DL models. The main contribution of this work is two-fold:
\begin{itemize}
    \item We implement a transformer based method that incorporates positional encoding, multi-head attention, layer normalization, and dropout techniques, and synthesizes the ABP waveform with a mean absolute error (MAE) of 3.01. Further, the MAE of SBP and DBP are 3.77 and 2.69, respectively. This method satisfies AAMI criterion, and achieves grade A, according to BHS criterion.  
    \item We implement a frequency-domain (FD) learning method where we first obtain the discrete cosine transform (DCT) coefficients of the PPG and ABP signals, and then learn a linear/non-linear (L/NL) regression between them. This method synthesizes the ABP waveform with an MAE of 4.23. Further, the MAE of SBP and DBP are 4.37 and 3.91, respectively. This method satisfies AAMI criterion, and achieves grade B, according to BHS criterion. 
\end{itemize}
We note that this work utilizes the UCI dataset (that contains data of patients in ICU), thus, the proposed methods do CLBP estimation under the condition when the subject is at rest.


{\bf Feasibility.} PPG and ABP signals, though of different origins, have a very similar morphology (because both capture the pressure exerted by the blood volume on arterial walls). In other words, the morphology of the two signals is tightly binded to each other, at the sub-cardiac cycle resolution. Furthermore, the recent advances in the generative artificial intelligence (AI) has made it possible to translate one biosignal to another, e.g., \cite{mehmood2023your} translates a PPG signal to an ECG signal, \cite{ibtehaz2022ppg2abp} translates a PPG signal to an ABP signal. This motivates us to propose purpose-built DL models for synthesis of the ABP waveform from a single-site PPG waveform.

{\bf Outline.} Section II provides some necessary background and summarizes selected related work. Section III provides compact details of the public UCI dataset that we have used, and the key data pre-processing steps that we have performed. Section IV describes the two deep learning models we have implemented for cuff-less arterial blood pressure waveform synthesis. Section V presents some selected results. Section VI concludes the paper.


\section{Background \& Related Work}

Before we summarize the selected related work, it is imperative to briefly introduce the so-called pulse wave analysis methods, which act as scaffolding for the design of most of the CLBP estimation methods.

\subsection{Background: Pulse wave analysis methods}

During the last decade, researchers have made extensive investigation of the physiological waveforms acquired in a non-invasive manner (e.g., ECG, PPG, PCG, etc.), under the umbrella term of pulse wave analysis (PWA) \cite{sola2019handbook}. The aim of PWA methods is to develop a number of indices to indirectly measure the blood pressure in a cuff-less manner using wearables (e.g., smart watches) and assistive gadgets (e.g., smartphones) \cite{korean_position_paper}. Some prominent indices (markers for blood pressure) include the following: pulse transit time (PTT), pulse arrival time (PAT), and pulse wave velocity (PWV). Before we delve into the discussion of the related work, it is imperative to formally define the three indices (BP surrogates) as follows: 
\begin{itemize}
    \item PTT is the time taken by a pulse pressure wave to travel from one point of an artery to another. PTT is inversely proportional to the blood pressure. A typical measurement system for PTT includes two PPG sensors placed at two different locations close to the artery \cite{7118672}, \cite{shriram2010ptt}. 
    \item PAT is the time delay between the R-peak in the ECG signal and the time of arrival of the corresponding (PPG) pulse at the peripheral site, during the same heart beat. In other words, PAT is the time taken by the pulse pressure wave to travel from the heart to the peripheral measurement site. PAT is related to PTT as follows: PAT = PTT + pre-ejection period \cite{sola2019handbook}. 
    \item PWV is the velocity of a pulse pressure wave that travels a distance $L$ from a proximal point to a distal point. PWV is related to PTT as follows: $PWV = \frac{L}{PTT}$ \cite{sola2019handbook}.  
\end{itemize}

Despite the good correlation of PTT, PAT and PWV with the blood pressure, these indices have a number of limitations as well, e.g., need for simultaneous and synchronous measurement of two waveforms from two different sites. Furthermore, there is a lack of consensus on translation of PTT/PAT/PWV values to SBP and DBP values. For example, one way to measure the SBP and DBP values through PTT and PWV is as follows:
\begin{equation}
    \text{SBP} = \alpha_{\text{s}} + \beta_{\text{s}} \cdot \text{PTT} + \gamma_{\text{s}} \cdot \text{PWV}
\end{equation}

\begin{equation}
    \text{DBP} = \alpha_{\text{d}} + \beta_{\text{d}} \cdot \text{PTT} + \gamma_{\text{d}} \cdot \text{PWV}
\end{equation}
where the specific coefficients $\alpha_{\text{s}}, \beta_{\text{s}}, \gamma_{\text{s}}, \alpha_{\text{d}}, \beta_{\text{d}}, \gamma_{\text{d}}$ need to be determined through empirical calibration with actual blood pressure measurements.
Thus, such translation methods have calibration issues, as one needs to re-celibrate on new datasets. In fact, there exist a number of equations to map PTT/PAT/PWV to SBP and DBP values, each with many unknowns that are subject-specific \cite{8513054}, \cite{sola2019handbook}, \cite{7118672}.

{\it This highlights the need for the design of novel single-site, calibration-free, non-parametric methods for cuff-less blood pressure estimation.}


\subsection{Related work}

Since the literature on CLBP estimation is quite vast, we could only summarize selected related works due to space constraints (see \cite{sola2019handbook}, \cite{7118672} and the references therein, which provide a more detailed treatment of the subject). The CLBP estimation methods are mostly PWA methods which are either two-sites/two-waveforms methods or single-site methods. More recently, researchers have extensively used various deep learning models for ABP waveform synthesis and CLBP estimation.

{\it Single-site-based methods:}
\cite{9145946} utilizes single-site PPG data collected from 150 subjects, computes PWV, and feeds it to a learning-based non-parametric regression method in order to estimate MABP, SBP and DBP. \cite{7590777} collects multi-wavelength, single-site PPG data from 10 subjects, computes the PTT using the IR-PPG and blue-PPG, in order to estimate the DBP and SBP. \cite{8267981} recruits 22 young healthy subjects to measure their ballistocardiogram (BCG) using the foot-based pressure sensor/force plate, in order to estimate the DBP (SBP) using I-J interval (J-K amplitude) of a BCG. \cite{ibtehaz2022ppg2abp} utilizes the UCI dataset and implements the well-known U-Net model to synthesize an ABP waveform from a single PPG waveform. 

{\it Two-sites-based methods:}
\cite{8438854} proposes a two-sites-based method for CLBP estimation that utilizes a PPG sensor at the fingertip and an impedance plethysmography (IPG) sensor at the wrist in order to compute the PTT which is then translated to SBP and DBP values. \cite{4463187} recruits 12 young healthy subjects, records their ECG and fingertip PPG data at multiple regular intervals after treadmill exercise, computes the PTT and translates it to the SBP and DBP values. \cite{kachuee2016cuffless} utilizes the ECG and PPG signals from MIMIC-II dataset, pre-processes them, extracts various physiological features, does dimensionality reduction, and implements a number of machine learning regression models to estimate SBP, MABP and DBP. 
\cite{4524030} adopts a rather different approach for CLBP estimation whereby authors utilize data from the PPG, pressure, and height sensors (accelerometers), exploiting the fact that movement of the arm changes the natural hydrostatic pressure.  

{\it Deep learning-based methods:}
\cite{9629687} utilizes the PPG and ABP signals corresponding to 20 subjects from the MIMIC II dataset, and implements a convolutional neural network (CNN) for feature extraction, followed by an long short-term memory (LSTM) network for CLBP prediction. \cite{mou2021cnn} considers 3 subjects from MIMIC II dataset, and utilizes a CNN-LSTM model for ABP waveform synthesis. Paviglianiti et al. \cite{paviglianiti2020noninvasive} utilize both PPG and ECG signals form the MIMIC dataset, and implement a number of DL models, i.e., ResNet, LSTM, WavNet, and more, for CLBP prediction. Brophy et al. \cite{brophy2021estimation} utilize PPG signals from the UCI dataset, implement a number of generative adversarial networks (GAN) models in order to realize a federated learning approach for ABP waveform synthesis. They evaluate the performance of their GAN-based federated learning model on a new unseen dataset, the University of Queensland vital signs dataset. Finally, \cite{8856706} utilizes a CNN, while \cite{8404255} implements a feedforward neural network for CLBP estimation.

\section{The UCI Dataset \& Data Pre-processing}

\subsection{The UCI public dataset on CLBP estimation}

We utilize the cuff-less blood pressure estimation dataset from the University of California Irvine (UCI) machine learning repository \cite{uci_cuff-less_blood_pressure_estimation}. This dataset is basically a subset of (and pre-processed version of) the MIMIC-II waveform dataset from Physionet databank, which is a large database of electronic health records of the patients from the intensive care unit of the Beth Israel Deaconess Medical Center in Boston, Massachusetts. This UCI dataset contains 12,000 records of (942 patients) that consists of the physiological waveforms carrying information about the vital signs, with length of each record varying between 8 sec and 10 min. Each record is sampled at 125 Hz and contains fingertip PPG, ECG (channel II), and invasive ABP signal (in mmHg). Further, we note that the dataset labels (i.e., DBP and SBP values) lie in the following range: (50$\le$DBP$\le$110) mmHg and (80$\le$SBP$\le$180) mmHg \cite{uci_cuff-less_blood_pressure_estimation}. In this work, we mainly focus on the PPG and ABP waveforms of first 209 subjects for CLBP estimation.

\subsection{Data Preprocessing} 

In order to prepare the data for our DL models, we implemented the following data preprocessing steps on the PPG and ABP signals. 
1) We began by removing out-of-band noise from the data using a Butterworth low-pass filter with a cut-off frequency of 10 Hz. 
2) We then proceeded to remove the residual artifacts manually (by discarding the anomalous samples). 
3) Next, we removed the baseline from the data by means of five-level Wavelet decomposition and reconstruction using db4 mother wavelet. 
4) Further, the process of ABP waveform synthesis from PPG waveform required beat-level synchronization between the PPG and ABP signals. To this end, we utilized time shifting (cross-correlation) approach to align the systolic peaks of the PPG signal $x[n]$ with the systolic peaks of the ABP signal $y[n]$:
\begin{equation}
R_{xy}(m) = \sum_{n=0}^{N-1} x[n] y[n-m]
\end{equation}
where \( m \) is the lag variable, and ranges from \(-N+1\) to \(N-1\).

\noindent
5) Inline with previous works \cite{shokouhmand2024mems,omer2024video}, we did segmentation (with a stride of 0) to construct a number of segments from each PPG and ABP recording, in order to increase the size of the dataset. Note that each segment is of duration 2 seconds (i.e., it consists of 2-3 cardiac cycles). 
6) Finally, we normalized the segmented data using z-score normalization: $z = \frac{q_s - \mu}{\sigma}$.
where \( z \) is the normalized data, \( x \) is the original data, \( \mu \) is the mean across the data segment, and \( \sigma \) is the standard deviation across the segment.

\section{Methodology}

Keeping in mind that the problem at hand---ABP waveform synthesis from PPG waveform---is a sequence-to-sequence translation (i.e., regression) problem, we have implemented two novel deep learning methods using Tensorflow and Keras frameworks in Python: 1) a purpose-built transformer model, 2) a frequency-domain linear/non-linear regression model. Once the ABP waveform is synthesized, we sample its maximum and minimum values which yield the systolic and diastolic blood pressure, respectively.

\subsection{Transformer model for ABP waveform synthesis}
The motivation for implementing a transformer model for ABP waveform synthesis from PPG waveform comes from the following: i) both the PPG and the ABP waveforms represent the time-series data, ii) each of the two time-series consists of an ordered set of data points that capture the temporal dynamics of the heart, i.e., the heart physiology, at different time-scales and on its own, iii) there is one-to-one correspondence (relationship) between the (ordered) data points (at the same instant) in the PPG and ABP time-series, due to the fact that they both capture the same phenomenon (of heart physiology), though independently. Keeping in mind all these strong analogies with the natural language processing (NLP) problems (e.g., language translation, paraphrasing, etc.), we solve the ABP waveform synthesis problem using the transformer model---a powerful tool for solving a range of NLP problems. 

Since we aim to solve a sequence-to-sequence translation problem, our purpose-built transformer model consists of an encoder-decoder pair. During the implementation phase, we tried several configurations of the transformer model, and saved the best performing configuration. 
Table \ref{table:arch_tx} summarizes the architecture of our purpose-built transformer. One can see that both the encoder and decoder utilize a dense layer, followed by positional encoding layers, followed by three transformer blocks. The outputs of the decoder and encoder blocks are fused by means of a multi-head cross attention mechanism. This is followed by a couple of dense layers at the end which produce the final output, i.e., ABP signal.



Note that each transformer block consists of the following layers: multi-head self-attention mechanism and feed-forward neural networks, with residual connections and layer normalization applied after each sub-layer. Further, it is imperative to explain the working of the following building blocks of our transformer model:

1) {\it Positional encoding:} It keeps the transformer informed about the order and position of data points (i.e., the notion of time) in each of the two (i.e., PPG and ABP) sequences. 
For a position \( pos \) and dimension \( i \):
\begin{equation}
PE_{(pos, 2i)} = \sin \left( \frac{pos}{10000^{\frac{2i}{d_{model}}}} \right) \quad (\text{for even indices})
\end{equation}

\begin{equation}
PE_{(pos, 2i+1)} = \cos \left( \frac{pos}{10000^{\frac{2i}{d_{model}}}} \right) \quad (\text{for odd indices})
\end{equation}
where $d_{model}$ is the dimension of the model (i.e., the size of the embedding vectors). 

2) {\it Multi-head self-attention:} As the name implies, it focuses on capturing relationship between every pair of data points (by means of attention scores) in both the PPG and ABP sequences, using multiple heads that attend to different parts of the PPG and ABP sequences in order to capture the short term and the long term dependencies within the PPG and ABP signals. Multi-head self-attention mechanism consists of the following key steps.
Let $X$ represent input embeddings of shape \((n, d_{\text{model}})\) with $n$ being the sequence length. First, the query, key, and value matrices $Q,K,V$ are obtained by linear projections of the input embeddings as follows: $Q = X W_Q$, $K = X W_K$, $V = X W_V$ where
$W_Q \in \mathbb{R}^{d_{\text{model}} \times d_k}, \quad W_K \in \mathbb{R}^{d_{\text{model}} \times d_k}, \quad W_V \in \mathbb{R}^{d_{\text{model}} \times d_v} $
are the three learned weight matrices, and $d_k,d_v$ represent the dimension of the key, value vectors, respectively.
This allows us to compute the scaled dot-product attention as follows:
\begin{equation}
\text{Attention}(Q, K, V) = \text{softmax}\left(\frac{QK^T}{\sqrt{d_k}}\right) V
\end{equation}
where softmax is an activation function defined as follows:
$\text{softmax}(\mathbf{z})_i = \frac{e^{z_i}}{\sum_{j=1}^{N} e^{z_j}}$
where \( \mathbf{z} \) is the input vector, \( z_i \) is the \( i \)-th element of the input vector \( \mathbf{z} \), \( N \) is the number of elements in the input vector \( \mathbf{z} \).
Finally, multi-head attention is computed as follows:
\begin{equation}
\text{MultiHead}(Q, K, V) = \text{Concat}(\text{head}_1, \ldots, \text{head}_h) W^O
\end{equation}
where $h$ is number of heads, and $W^O$ is the output projection matrix. Further,
\begin{equation}
\text{head}_i = \text{Attention}(Q_i=XW_{Q_i}, K_i=XW_{K_i}, V_i=XW_{V_i})
\end{equation}
where $W_{Q_i}, W_{K_i}, W_{V_i}$ are the query, key, value projection matrices for the i-th head.

\begin{table}[]
\centering
\begin{adjustbox}{max width=\textwidth}
\begin{tabular}{|>{\raggedright\arraybackslash}m{2.2cm}|>{\raggedright\arraybackslash}m{1.9cm}|>{\raggedright\arraybackslash}m{0.9cm}|>{\raggedright\arraybackslash}m{2cm}|}
\hline
\textbf{Layer type}          & \textbf{Output Shape} & \textbf{No. of Param } & \textbf{Connected to} \\
\hline
Input (e)         & (None, 250, 1)        & 0                 & -                   \\
\hline
Input (d)        & (None, 250, 1)        & 0                 & -                   \\
\hline
Dense (e)              & (None, 250, 64)       & 128               & Input (e)  \\
\hline
Dense (d)              & (None, 250, 64)       & 128               & Input (d)  \\
\hline
Pos. Enc. (e) & (None, 250, 64) & 0 & Dense (e) \\
\hline
Pos. Enc. (d) & (None, 250, 64) & 0 & Dense (d) \\
\hline
Tx Block 1 (e) & (None, 250, 64) & 74944 & Pos. Enc. (e) \\
\hline
Tx Block 1 (d) & (None, 250, 64) & 74944 & Pos. Enc. (d) \\
\hline
Tx Block 2 (e) & (None, 250, 64) & 74944 & Tx Block 1 (e) \\
\hline
Tx Block 2 (d) & (None, 250, 64) & 74944 & Tx Block 1 (d) \\
\hline
Tx Block 3 (e) & (None, 250, 64) & 74944 & Tx Block 2 (e) \\
\hline
Tx Block 3 (d) & (None, 250, 64) & 74944 & Tx Block 2 (d) \\
\hline
MultiHead Attention & (None, 250, 64) & 66368 & Tx Block 3 (e), Tx Block 3 (d) \\
\hline
Dense 1              & (None, 250, 64)       & 4160              & MultiHead Attention \\
\hline
Dense 2              & (None, 250, 1)        & 65                & Dense 1  \\
\hline
\end{tabular}
\end{adjustbox}
\caption{Architecture of our purpose-built transformer model for ABP waveform synthesis. (e) implies an encoder layer, while (d) implies a decoder layer. Finally, Tx is the short for transformer. }
\label{table:arch_tx}
\end{table}



\subsection{Frequency-domain linear/non-linear regression model for ABP waveform synthesis}

In this method, we follow a two-step approach for ABP waveform synthesis. During the training phase, the two steps are implemented as follows: 
\begin{itemize}
    \item {\it Frequency-domain representation of pre-processed data:} Firstly, we obtain the frequency-domain representation of the pre-processed data, for both signals of interest, i.e., PPG and ABP signals. The motivation for this comes from the fact that both PPG and ABP signals are quasi-periodic due to the rhythmic activity of the heart. Specifically, we consider one PPG-ABP segment-pair corresponding to 2-3 cardiac cycles of duration $Q'$ samples, at a time. We then compute the discrete cosine transform (DCT) (type-II) of the PPG segment \( x_n \) of length \( Q' \) in the single PPG-ABP segment-pair:
\begin{equation}
X_k = \sum_{n=0}^{Q'-1} x_n \cos \left( \frac{\pi}{Q'} \left( n + \frac{1}{2} \right) k \right) \quad k = 0, \ldots, Q'-1
\end{equation}
Similarly, we compute the DCT of the corresponding ABP segment in the single PPG-ABP segment-pair. 
    We observe that the PPG and ABP signals are sparse in frequency domain (with only a few low-frequency components with significant energy). This allows us to have a compact representation of each of the two segments in the given PPG-ABP segment-pair by retaining first $Q_X$ ($Q_Y$) significant DCT coefficients of the PPG (ABP) segment only. This in turn helps in reducing the computationally complexity of our linear/non-linear (L/NL) regression model by reducing the dimension of the feature vector it takes as input. Note that we zero-pad both DCT vectors of size $Q_X$ and $Q_Y$ to resize them to $Q$ in order to do the L/NL regression in the next step (where $Q$ represents the duration of a segment, in samples). 
    \item {\it Linear/non-linear regression:} Secondly, we learn the linear/non-linear regression between the size-$Q$ DCT coefficients vector $\mathbf{x}$ (that represents the most important features of the PPG signal) and the size-$Q$ DCT coefficients vector $\mathbf{y}$ (that represents the most important features of the ABP signal). For linear regression, we have implemented ridge regression\footnote{Ridge regression, also known as Tikhonov regularization, is basically linear regression but with an additional term for $L2$ regularization that helps prevent overfitting and stabilize the regression coefficients.}. 
The ridge regression solution is given by:
\begin{equation}
\mathbf{w} = (\mathbf{X}^T \mathbf{X} + \lambda \mathbf{I})^{-1} \mathbf{X}^T \mathbf{y}
\end{equation}
where
    \( \mathbf{X} \) is a \( Q \times Q \) matrix of DCT vectors $\mathbf{x}$ of PPG segments,
    \( \mathbf{y} \) is a \( Q \times 1 \) vector of DCT coefficients of ABP segment,
    \( \lambda \) is a scalar regularization parameter,
    \( \mathbf{I} \) is the identity matrix of size \( Q \times Q \),
    $\mathbf{w}$ is the vector of regression coefficients.
\end{itemize}




Next, the testing phase. We compute DCT (type-II) of a previously unseen size-$Q'$ PPG segment, resize it to $Q_X$ and then $Q$. We then feed the size-$Q$ DCT vector (of PPG) to the L/NL frequency-domain regression model which outputs a size-$Q$ DCT vector (of ABP). Then, the ABP waveform $x_n$ is synthesized by taking the inverse DCT (I-DCT) of the output $X_k$ of the L/NL regression model:
\begin{equation}
x_n = \sum_{k=0}^{Q-1} \alpha_k X_k \cos \left( \frac{\pi}{Q} \left( n + \frac{1}{2} \right) k \right) \quad n = 0, \ldots, Q-1
\end{equation}
where the coefficient \( \alpha_k \) is defined as:
\begin{equation}
\alpha_k = 
\begin{cases}
\sqrt{\frac{1}{Q}} & \text{if} \ k = 0 \\
\sqrt{\frac{2}{Q}} & \text{if} \ k > 0
\end{cases}
\end{equation}

\begin{figure}
  \centering
  \includegraphics[width=8.5cm]{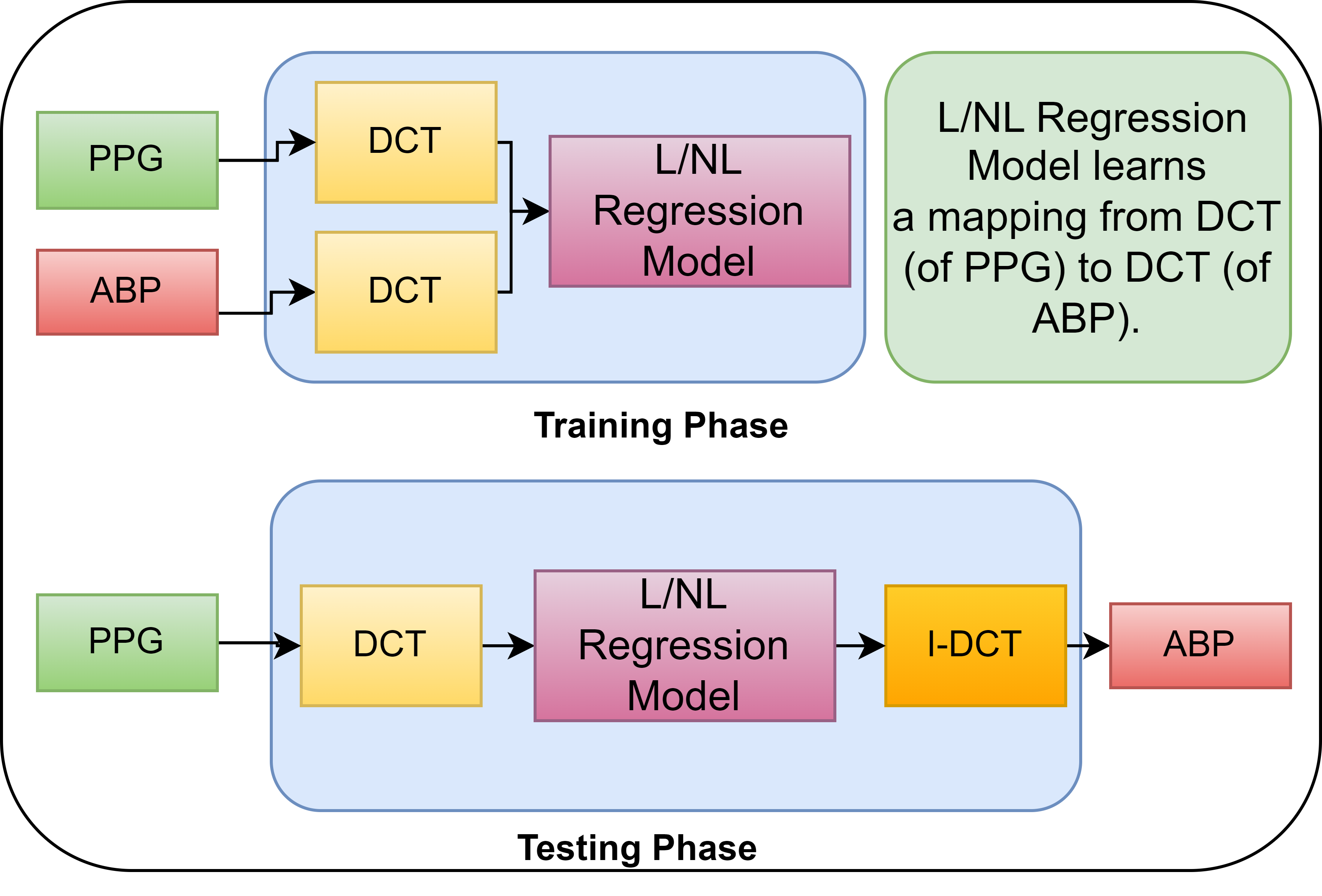}
  \caption{Architecture of FD L/NL regression model for ABP waveform synthesis.}
  \label{Figure:DCTmodel}
\end{figure}

Fig. \ref{Figure:DCTmodel} provides a pictorial overview of the proposed FD L/NL regression model.

\section{Results and Discussion}

{\it Performance Metrics:}
We have utilized the following two metrics to evaluate the performance of our DL models for CLBP estimation: mean absolute error (MAE), and root mean squared error (RMSE), where MAE = $\frac{1}{N}\sum_{i=1}^{N}|\hat{y}_i-y_i|$, and $\text{RMSE}(y, \hat{y}) = \sqrt{\frac{\sum_{i=1}^{N} (y_i - \hat{y}_i)^2}{N}}$. Here, $N$ represents the batch size, $y$ represents the reference value/ground truth, and $\hat{y}$ represents the model prediction. 

\subsection{Performance evaluation of Transformer model}

{\it Setting of hyperparameters:}
We have a total of 32,052 samples/examples where each sample has duration 4 seconds (or, 1,000 data points). We set segment size to 2 seconds, i.e., 250 points. This changes the data shape from (32,052, 1,000) to (128,208, 250). 
For training and evaluation of transformer model, we utilized a 72:8:20 train-validation-test split. 
We set num-heads = 4, ff-dim = 64, num-transformer-blocks = 3, embed-dim = 64, epochs=3, batch-size=128, and optimizer=Adam. 
For each transformer block, we set epsilon to 1e-6 for normalization layer, and dropout ratio to 0.1.   

\begin{figure*}
  \centering
  \includegraphics[width=0.9\textwidth]{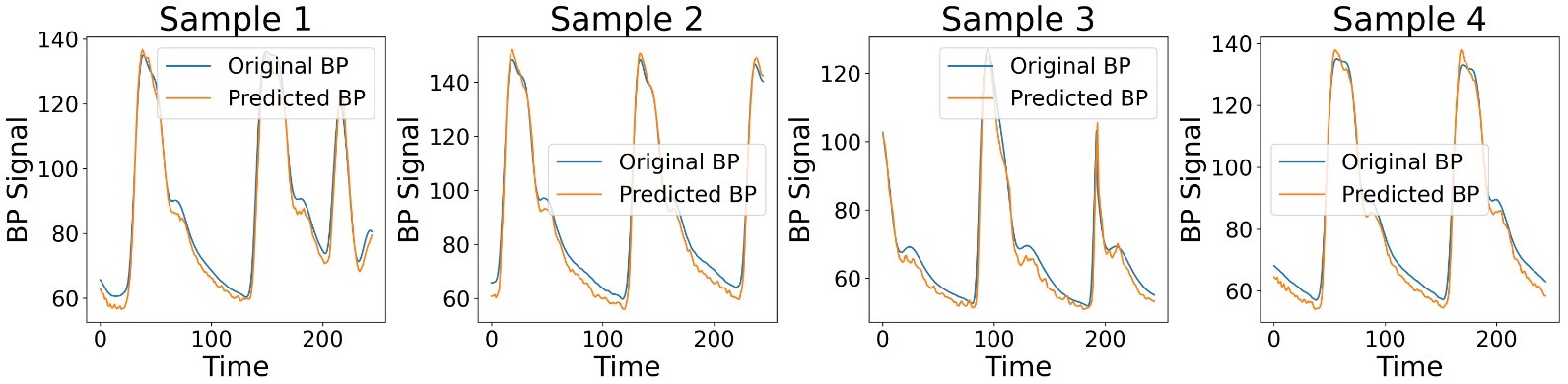}
  \caption{Four examples of ABP waveform synthesis by our transformer model.}
  \label{fig:ppgtoabpbytrans}
\end{figure*}

{\it Qualitative results:}
Fig. \ref{fig:ppgtoabpbytrans} shows four examples where the transformer model synthesizes the ABP signal from a PPG signal. It is clear that the morphology of the synthetic ABP waveforms constructed by the transformer model is quite similar to the morphology of the reference ABP waveforms. 

{\it Quantitative results:} 
The transformer model achieves an MAE of 3.01 for ABP waveform synthesis, an MAE of 3.77 mmHg for SBP estimation, and an MAE of 2.69 mmHg for DBP estimation. 


{\it Computational Complexity of Transformer Model:}
Recall that our transformer model consists of dense layers, positional encoding, transformer blocks, and multi-head attention block. We calculate its computational complexity by analyzing each layer and summing up the contributions. 
\begin{itemize}
\item {Dense layers:}
For a dense layer with input dimension \( n \) and output dimension \( m \), the number of parameters is \( n \times m + m \), and the computational complexity is \( O(n \times m) \). Our purpose-built transformer consists of 4 dense layers (see Table \ref{table:arch_tx}). The following holds for the first two dense layers: $\text{input} \ (250, 1) \ \rightarrow \text{output} \ (250, 64)$, $\text{parameters:} \ 1 \times 64 + 64 = 128$, $\text{complexity:} \ O(250 \times 64) = O(16000)$. Next, for the 3rd dense layer, $\text{input} \ (250, 64) \ \rightarrow \text{output} \ (250, 64)$, $\text{parameters:} \ 64 \times 64 + 64 = 4160$, $\text{complexity:} \ O(250 \times 64^2) = O(1,024,000)$. Finally, for the fourth dense layer, $\text{input} \ (250, 64) \ \rightarrow \text{output} \ (250, 1)$, $\text{parameters:} \ 64 \times 1 + 1 = 65$, $\text{complexity:} \ O(250 \times 64) = O(16000)$.
\item {Positional encoding:}
Positional encoding does not have parameters, and is computationally negligible.
\item {Transformer blocks:}
A transformer block consists of multi-head self-attention and feed-forward network.
Self-attention complexity: $O(N \times d^2)$.
Feed-forward network complexity: $O(N \times d^2)$.
In our work, $N = 250, \ d = 64$.
This implies that the complexity of self-attention is:
$O(250 \times 64^2) = O(1,024,000)$.
The complexity of feed-forward network is:
$O(250 \times 64^2) = O(1,024,000)$.
Thus, total complexity for one transformer block is:
$O(2 \times 1,024,000) = O(2,048,000)$. Finally, total complexity for six transformer blocks is:
$6 \times 2,048,000 = 12,288,000$.

\item {Multi-head attention:}
This block consists of multiple self-attention mechanisms. In our model, \( h = 4 \). Thus, 
$\text{multi-head attention complexity is:} \; O(4 \times 1,024,000) = O(4,096,000)$.
\item {Overall complexity:}
We sum up the complexities of four dense layers, six transformer blocks, and multi-head attention block to get the total computational complexity of our transformer model:
$O(1,072,000 + 12,288,000 + 8,192,000) = O(21,552,000)$.

\end{itemize}

\subsection{Performance evaluation of FD L/NL Regression model}

{\it Setting of hyperparameters:}
For ridge regression, we vary the regularization parameter $\alpha$ in the range $1e-3$ to $1e3$. The optimal $\alpha$ turned out to be $1e0$. The rest of the hyperparameters remain the same.

\begin{figure}
  \centering
  \includegraphics[width=9cm, height=7cm]{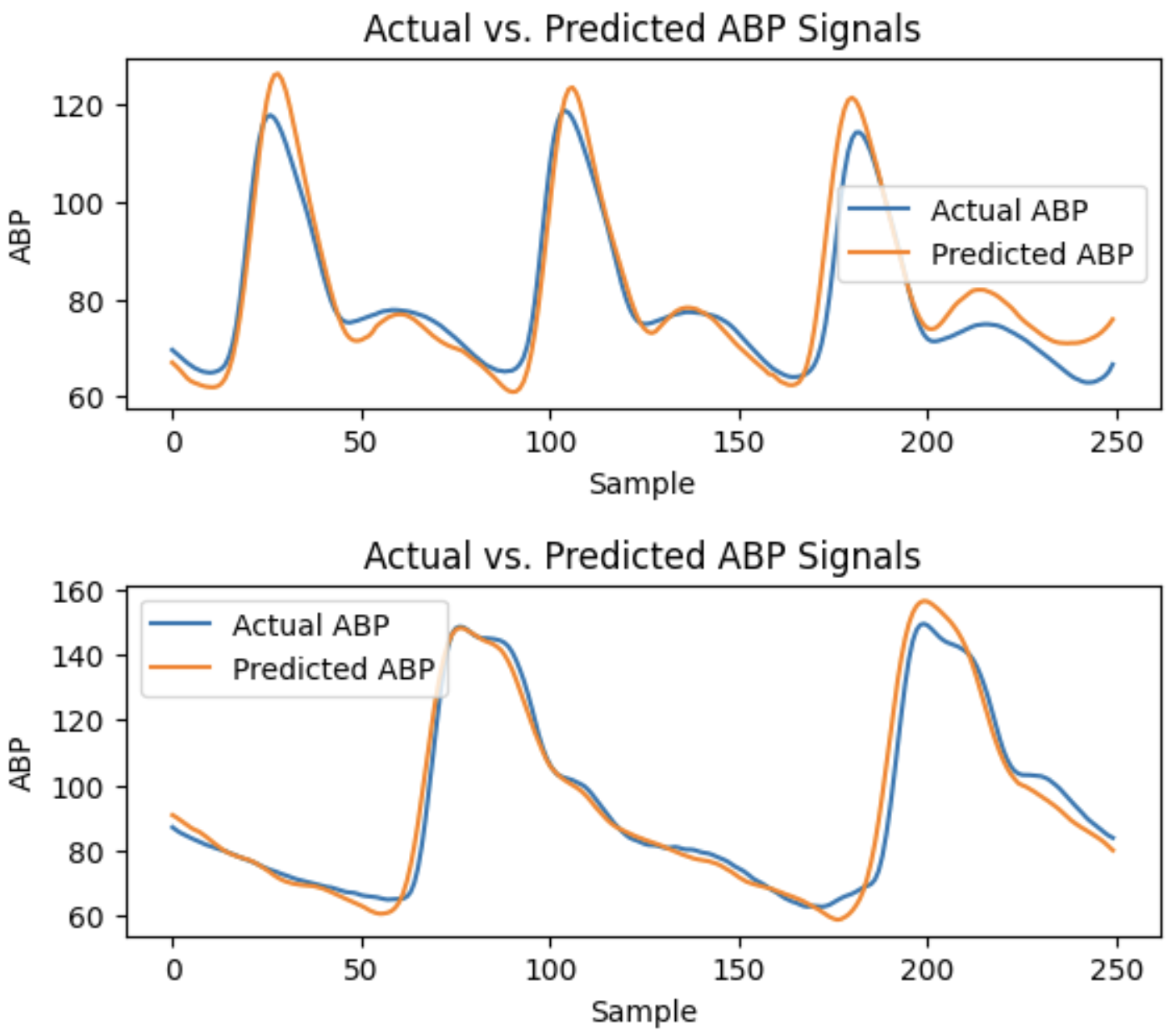}
  \caption{Two examples of ABP waveform synthesis by our FD L/NL regression model.}
  \label{fig:ppgtoabpbydct}
\end{figure}

{\it Qualitative results:} 
Fig. \ref{fig:ppgtoabpbydct} presents two examples to showcase the quality of the ABP signal synthesized by the FD L/NL regression model. It is evident that the morphology of the two synthetic ABP waveforms produced by the FD L/NL regression model (on the right) and the morphology of the reference ABP waveforms (on the left) are highly similar. 


{\it Quantitative results:} 
The FD L/NL model achieves an MAE of 4.23 for ABP waveform synthesis, an MAE of 4.37 mmHg for SBP estimation, and an MAE of 3.91 mmHg for DBP estimation.

{\it Computational Complexity of the FD L/NL Model:}
The computational complexity of the proposed FD L/NL model that involves taking the DCT of a PPG signal, computing the ridge regression of this DCT with the DCT of ABP signal, and then taking the I-DCT is given by:
\begin{itemize}
\item {DCT complexity:}
The complexity of computing the DCT of a signal of length \( N \) is: $O(N \log N)$.
\item {Ridge regression complexity:}
Assuming we have \( m \) features and \( n \) samples, the complexity of ridge regression is: $O(m^2 n + m^3)$.
\item {I-DCT complexity:}
The complexity of computing the inverse DCT of a signal of length \( N \) is: $O(N \log N)$.
\item {Combined complexity:}
Combining the complexities of DCT, ridge regression, and IDCT, we get: $2 \times O(N \log N) + O(m^2 N + m^3)$.
In Big-O notation, constants are omitted, so the final computational complexity is: $O(N \log N) + O(m^2 N + m^3)$.
\end{itemize}

\subsection{Performance comparison with related work}

Table \ref{tab:sota} provides a detailed performance comparison of our work with the selected related works by providing a quick summary of different deep learning models used for CLBP estimation, their corresponding datasets, MAE performance achieved by each model, and grade obtained according to the BHS and AAMI criteria. Table \ref{tab:sota} highlights the variability in the performance of the related works, which is in turn due to the change of data distribution across datasets, variable sizes of the datasets, model architectures, etc. Our first approach, i.e., the transformer model achieves an MAE of 3.01 for ABP waveform synthesis, an MAE of 3.77 mmHg for SBP estimation, and an MAE of 2.69 mmHg for DBP estimation. Further, it fulfills AAMI criterion and obtains grade A by BHS criterion.  
Our 2nd approach, i.e., the FD L/NL regression model achieves an MAE of 4.23 for ABP waveform synthesis, an MAE of 4.37 mmHg for SBP estimation, and an MAE of 3.91 mmHg for DBP estimation. It also fulfills AAMI criterion and obtains grade B by BHS criterion.  
We note that nearly all the works suffer from the issue of lack of generalization, i.e., as the number of subjects increases, the MAE of the SBP and DBP increases, while the BHS grade tends to fall \cite{koparir2024cuffless}, \cite{figini2022improving}.

\begin{table*}[h]
    \centering
    \caption{Performance comparison of our work with the state-of-the-art CLBP estimation methods}
    \begin{tabular}{@{}lccccccc@{}}
        \toprule
        \textbf{Methodology} & \textbf{MAE (SBP)} & \textbf{MAE (DBP)} & \textbf{Dataset} & \textbf{Subjects} & \textbf{BHS Grade} & \textbf{AAMI Grade} & \textbf{Reference} \\ \midrule
        XGBoost & 5.45 mmHg & 3.88 mmHg & custom SPG dataset & 35 & - & - & \cite{shokouhmand2024mems} \\
        Waveform-based ANN-LSTM & 2.34 mmHg & 1.77 mmHg & PPG-DaLiA & 15 & A & Pass & \cite{tanveer2019cuffless} \\
        CNN with transfer learning & 5.63 mmHg & 2.82 mmHg & MIMIC-II Database & 942 & A & Pass & \cite{koparir2024cuffless} \\
        DCT and Regression-Based Methods & 4.02 mmHg & 2.78 mmHg & MIMIC Database & 200 & - & Pass & \cite{figini2022improving} \\
        Stacked Deformable CNN & 4.86 mmHg & 4.42 mmHg & Custom Clinical Dataset & 30 & A & - & \cite{qiu2024non} \\ 
        CNN-LSTM & 3.7 mmHg & 2.02 mmHg & MIMIC-II Dataset & 20 & A & - & \cite{9629687} \\
        {\bf This work (transformer method)} & {\bf 3.77 mmHg} & {\bf 2.69 mmHg} & {\bf UCI Database} & {\bf 209} & {\bf A} & \bf{Pass} & -  \\
        {\bf This work (FD L/NL method)} & {\bf 4.37 mmHg} & {\bf 3.91 mmHg} & {\bf UCI Database} & {\bf 209} & {\bf B} & {\bf Pass} & - \\
        \bottomrule
    \end{tabular}
    \label{tab:sota}
\end{table*}

\section{Conclusion}

We evaluated two novel purpose-built DL models on public UCI dataset for CLBP estimation using a single-site PPG signal. The transformer model (the FD L/NL regression model) achieved an MAE of 3.77 and 2.69 (4.37 and 3.91), for DBP and SBP, respectively. Both models fulfilled the AAMI criterion. Finally, the transformer model (FD L/NL model) achieved grade A (grade B), according to the BHS criterion. Thus, this work contributes to the efforts which aim to do SBP and DBP estimation using a single bio-signal obtained from a single site, e.g., through a smartwatch, smartband, etc.

We note that the UCI dataset consists of data from patients in intensive care units, who are often under the effects of drugs and other medical interventions. Thus, it is imperative to look into other populations, such as healthy adults doing various physical activities. Therefore, the future work will look into the subject-specific training and evaluation of our AI models, and evaluation and fine-tuning of our AI models on other diverse datasets to assess and enhance their generalization capability. 


\bibliography{sources}
\bibliographystyle{ieeetr}
\end{document}